\documentclass[11pt,a4paper]{article}
 
\pdfoutput=1 

\usepackage{graphicx}
\usepackage{epstopdf}
\usepackage{jcappub}
\usepackage{amsmath,amsthm,latexsym,amssymb,amsfonts,epsfig}
\usepackage{hyperref}
\usepackage{psfrag}
\usepackage[cp1250]{inputenc}
\usepackage[usenames,dvipsnames]{xcolor}

\numberwithin{equation}{section}

\title{Constant roll warm inflation in high dissipative regime}

\author[a]{Vahid Kamali}
\author[b]{Micha{\l} Artymowski}
\author[c]{Mohammad Reza Setare}
\affiliation[a]{Department of Physics, McGill University, Montreal, QC, H3A 2T8,
Canada.\\
Department of Physics, Bu-Ali Sina (Avicenna) University, Hamedan 65178,
016016, Iran.\\
School of Physics,
Institute for Research in Fundamental Sciences (IPM),
19538-33511, Tehran, Iran}
\affiliation[b]{Institute of Theoretical Physics, Faculty of Physics, University of Warsaw, ul. Pasteura 5, 02-093 Warsaw, Poland}
\affiliation[c]{Department of Science, Campus of Bijar, University of Kurdistan, Bijar, Iran}
\emailAdd{vkamali@ipm.ir}
\emailAdd{michal.artymowski@fuw.edu.pl}
\emailAdd{rezakord@ipm.ir}

\abstract{Constant-roll warm inflation is introduced in this work. A novel approach to finding an exact solution for Friedman equations {in the constant-roll framework} is presented for cold inflation and is extended to warm inflation with the constant dissipative parameter $Q=\frac{\Gamma}{3H}$. The evolution of the primordial inhomogeneities of a scalar field in a thermal bath is also studied. The $1\sigma$ consistency between the theoretical predictions of the model and observational constraints has been proven for a range of $Q$ and $\beta=-\frac{\ddot{\phi}}{3H\phi}$ (constant rate of inflaton roll). In addition, we briefly investigate the possible enhancement of super-horizon perturbations beyond the slow-roll approximation.}


\begin{document}

\maketitle

\section{Introduction}

The cosmic inflation \cite{Starobinsky:1980te,Lyth:1998xn,Sato:1980yn} is a hypothetical, but well-motivated period of the accelerated expansion of the early Universe. Inflation solves problems of the horizon, curvature and monopoles \cite{Guth:1980zm}, which appear in the big bang cosmology. Inflation can also be responsible for the generation of the primordial inhomogeneities \cite{Akrami:2018odb, Planck:2013jfk}, which are the seeds of the large scale structure of the Universe. One of the most popular assumptions made in the inflationary paradigm is that the inflaton $\phi$ (i.e. the field which generates inflation) should evolve so slowly, that in its  equation of motion
\begin{equation}
\ddot{\phi} + 3H\dot{\phi} + V_\phi = 0 \,  \label{eq:EOMCold}
\end{equation}
one can assume that $\ddot{\phi} \ll V_\phi, 3H\dot{\phi}$, where $V_\phi = \frac{dV}{d\phi}$. Such a slow evolution of $\phi$ is usually equivalent to the following assumptions about the flatness of the potential
\begin{equation}
\epsilon = -\frac{\dot{H}}{H^2} \simeq \frac{M_p^2}{2}\left(\frac{V_\phi}{V}\right)^2 \ll 1 \, \qquad |\eta| = \left|\frac{\ddot{\phi}}{H\dot{\phi}}\right| \simeq \left|M_p^2 \frac{V_{\phi\phi}}{V}\right| \ll 1 \, , \label{eq:SRConditionsCold}
\end{equation}
where $\epsilon$ and $\eta$ are the slow-roll parameters. As noted in \cite{Martin:2012pe,Motohashi:2014ppa} one can obtain a quasi de-Sitter expansion of the Universe beyond the slow-roll regime. For instance for locally flat potential one can obtain $V_\phi  \ll \ddot{\phi} \Rightarrow \ddot{\phi} + 3H\dot{\phi \simeq 0}$, which leads to $|\eta| \sim 3$. Cases of the slow-roll and beyond the slow-roll evolution can be characterized by the condition $\ddot{\phi} + 3\beta H \dot{\phi} = 0$, where {$\beta$ takes values:} $\beta = 0$ (ultra slow-roll case) or $\beta=1$ (beyond slow-roll). The idea of the constant roll inflation is to take a continuous spectrum of $\beta$ \cite{Motohashi:2014ppa,Martin:2012pe,Motohashi:2017aob,Motohashi:2017gqb,Odintsov:2017qpp,Anguelova:2017djf}, which can also be realized in theories of modified gravity \cite{Motohashi:2017vdc,Nojiri:2017qvx,Karam:2017rpw}. {In the constant-roll approach one assumes that both \eqref{eq:EOMCold} and $\ddot{\phi} = -3\beta H\dot{\phi}$ are satisfied, which allows to reconstruct a scalar potential, that gives the constant-roll solution.}
\\*

Another theory of inflation, which assumes non-standard form of the cosmic friction term in the equation of motion is the warm inflation \cite{Berera:1995ie, Graham:2009bf, BuenoSanchez:2008nc, BasteroGil:2009ec, Berera:2002sp, Berera:2018tfc, Kamali:2019ppi, Setare:2012fg, Setare:2013dd, Setare:2013ula, Setare:2013qfa,Setare:2013dra,Setare:2013kja, Setare:2014oka,Setare:2014qea,Setare:2014uja,Setare:2014gya, Setare:2015cta, Kamali:2015yya, Kamali:2015dfl, Kamali:2016frd, Kamali:2017zgg, Basilakos:2017bol,Kamali:2017nwe, Motaharfar:2018zyb, Kamali:2018ylz}, for which one considers the dissipation of the energy density of inflaton  into relativistic degrees of freedom. This effect reheats the Universe through the whole period of inflation, which leads to the warm inflationary Universe. In this work, we look for constant-roll solutions in the warm inflationary scenarios. We also investigate the evolution of primordial inhomogeneities and look for a possibility of super-horizon enhancement of curvature perturbation \cite{Motohashi:2017kbs}. 

{The structure of this paper goes as follows: In Sec. \ref{sec:Cold} we redo the well-known work on constant-roll inflation solving the system as a function of e-folds and scalar field. In Sec. \ref{sec:Warm} the analysis is extended into the system with a constant-rolling scalar field, which dissipates into radiation. In Sec. \ref{sec:Pert} we investigate the primordial inhomogeneities in the warm inflation and we show that they are fully consistent with the Planck/Bicep data. Finally, we summarize in Sec. \ref{sec:Conclusions}}

\section{Constant-roll inflation} \label{sec:Cold}

Let us assume that the metric of the early Universe is a flat FRW metric and that the Universe is filled by a homogeneous scalar field $\phi$. In such a case one finds
\begin{eqnarray}
3M_p^2H^2 = \rho = \frac{1}{2}\dot{\phi}^2 + V \, , \label{eq:Fried1} \\
-2M_p^2 \dot{H} = \rho+p = \dot{\phi}^2 \, , \label{eq:Fried2} 
\end{eqnarray}
where $V_\phi =\frac{dV}{d\phi}$. To obtain constant roll inflation we impose a constant roll equation of motion, namely
\begin{equation}
\ddot{\phi} + 3\beta H\dot{\phi} = 0 \, , \label{eq:CR}
\end{equation}
which indicates, that
\begin{equation}
\dot{\phi} = \pm \dot{\phi}_0 a^{-3\beta} \, , \label{eq:phipCR}
\end{equation}
where $\dot{\phi}_0$ is a value of $\dot{\phi}$ at some initial moment $t_i$. We also assume that $a(t_i) = 1$. Number of e-folds is defined as
\begin{equation}
N = \log\frac{a}{a_i} = \log a \qquad \Rightarrow \qquad a = e^N \, ,
\end{equation}
which gives \footnote{A similar way of solving the equations of motion of the constant-roll inflation for $\beta = 1$ was presented in \cite{Firouzjahi:2018vet}.}
\begin{equation}
\dot{\phi} = \pm \dot{\phi}_0 e^{-3\beta N} \, .\label{eq:phiofN}
\end{equation}
{As we will show, for every value of $\beta$ one can find $\phi$, $V$, and $H$ as a function of $N$, obtain $N(\phi)$ and finally reconstruct the potential of the theory. Note that} one can re-write the LHS of the Eq. (\ref{eq:Fried2}) into
\begin{equation}
-2M_p^2\dot{H} = -2M_p^2 \frac{dH}{dN}\frac{dN}{dt} = -2M_p^2 H_N H = -M_p^2\frac{d}{dN}(H^2) \, ,
\end{equation}
which gives
\begin{equation}
-M_p^2\frac{d}{dN}(H^2) = \dot{\phi}_0^2 e^{-6\beta N} \qquad \Rightarrow \qquad 3M_p^2H^2 = \frac{\dot{\phi}_0^2}{2\beta}a^{-6\beta} + V_0 \, , \label{eq:Fried1SolCold}
\end{equation}
where $V_0$ is a constant of integration. One can interpret $V_0$ as a constant part of the inflationary potential or as a cosmological constant. In those two cases the value of $V_0$ could be:
\begin{itemize}

\item[1)] Assuming that $V(\phi)$ determines all of the evolution of the inflaton (including the graceful exit, reheating, etc.) the $V_0$ should be interpreted as a late-time cosmological constant and one should assume $V_0 \sim \rho_{DE} \ll V$ during inflation. This comes from the fact that $\rho_{DE} \sim 10^{-120} M_p^4$, which compared to the scale of inflation is negligible. 

\item[2)] The other approach is to assume that the constant-roll phase is just a part of the whole cosmic inflation and therefore the $V(\phi)$ approximates the true potential only in the vicinity of the constant-roll phase. In such a case the $V_0$ term may dominate over the other terms in the potential and one can expect $V_0$ to be {of order of} GUT scale. 

\end{itemize}

The RHS of the first Friedmann equation consists of the kinetic and potential term, namely
\begin{equation}
\rho = \frac{1}{2}\dot{\phi}^2 + V = \frac{1}{2}\dot{\phi}_0^2 a^{-6\beta} + V \, . \label{eq:RhoCold}
\end{equation}
Thus, from Eqs. (\ref{eq:Fried1SolCold},\ref{eq:RhoCold}) one finds
\begin{equation}
V = V_0 + \frac{1-\beta}{2\beta} \dot{\phi}_0^2 e^{-6\beta N} \, . \label{eq:VCold}
\end{equation}
In order to obtain $V > 0$ for all values of $N$ one requires $0<\beta \leq 1$. This constraint on $\beta$ has additional motivation - for $\beta > 1$ one obtains $V_N > 0$, which means, that the field moves uphill, while the Universe grows. Such a solution is possible in scalar-tensor theories \cite{Jinno:2017jxc,Jinno:2017lun,Artymowski:2018ewb}, but impossible in GR framework. For $\beta = 1$ one finds $V = V_0 = const$, which corresponds to the Universe filled with a massless homogeneous scalar field and a cosmological constant. This is a strongest deviation from the slow-roll approximation one can obtain in a de-Sitter Universe. It leads to big values of the $\eta$ parameter (since $\eta = \ddot{\phi}/H\dot{\phi} = -3\beta$) and to the growth of the super-horizon fluctuations \cite{Motohashi:2014ppa}. The same result as in (\ref{eq:VCold}) can be obtained by combining Eqs. (\ref{eq:EOMCold}) and (\ref{eq:CR}), which gives
\begin{equation}
3H(\beta - 1)\dot{\phi} = V_\phi \, . \label{eq:EOMCRCold}
\end{equation}
Note that
\begin{equation}
\frac{dV}{d\phi} = \frac{dV}{dN}\frac{dN}{dt}\frac{dt}{d\phi} = V_N \frac{H}{\dot{\phi}} \, ,
\end{equation}
which together with Eqs. (\ref{eq:phipCR},\ref{eq:EOMCRCold}) gives
\begin{equation}
V_N = 3(\beta - 1)\dot{\phi}_0^2 e^{-6\beta N} \qquad \Rightarrow \qquad V =  V_0 + \frac{1-\beta}{2\beta} \dot{\phi}_0^2 e^{-6\beta N} \, ,
\end{equation}
which is fully consistent with the result obtained in the Eq. (\ref{eq:VCold}). To obtain a potential as a function of field one needs to find $N = N(\phi)$. This can be done in two ways. First of all, one can use the fact that
\begin{equation}
\dot{\phi} = \frac{d\phi}{dN}\frac{dN}{dt} = H \phi_N = \pm \dot{\phi}_0 e^{-3\beta N} \, , \label{eq:phiNCondition}
\end{equation}
which, from Eqs. (\ref{eq:Fried1},\ref{eq:phipCR},\ref{eq:RhoCold}) gives
\begin{equation}
\phi(N) = \pm \sqrt{3} \, \dot{\phi}_0 M_p \int \frac{e^{-3\beta N} dN}{\sqrt{\frac{\dot{\phi}_0^2}{2\beta}e^{-6\beta N} + V_0}} \, . \label{eq:phiNCold}
\end{equation}
On the other hand one can use a fact that
\begin{equation}
3H (\beta - 1)\dot{\phi} = \frac{H}{\dot{\phi}} V_N \quad \Rightarrow \quad 3H^2(\beta - 1)\phi_N = \frac{V_N}{\phi_N} \quad \Rightarrow \quad \phi = \pm \frac{1}{\sqrt{\beta-1}}\int \sqrt{\frac{V_N}{3H^2}}dN \, , \label{eq:VCold2}
\end{equation}
which gives the same result as the Eq. (\ref{eq:phiNCold}). For $V_0 = 0$ one finds the following solution of the Eq. (\ref{eq:phiNCold})
\begin{equation}
\phi = \pm M_p \sqrt{6\beta} + \phi_0 \qquad \Rightarrow \qquad N = \pm \frac{\phi - \phi_0}{M_p\sqrt{6\beta}} \, , \label{eq:phiLinearCold}
\end{equation}
which gives
\begin{equation}
V = \frac{1-\beta}{2\beta} \dot{\phi}_0^2\exp\left(\mp \frac{\sqrt{6\beta}}{M_p}(\phi-\phi_0)\right) \, . \label{eq:VnoV0Cold}
\end{equation}
{This form of the potential has been presented in the Eq.(11) of Ref.\cite{Motohashi:2014ppa}, where $C_1=0$ or $C_2=0$.} One can explicitly see that the positivity of the potential requires $0<\beta \leq 1$. For $V_0=0$ one finds
\begin{equation}
\epsilon = -\frac{\dot{H}}{H^2} = 3\beta \, , \label{eq:EpsilonCold}
\end{equation}
which gives the inflationary solution only for $\beta \ll 1$. Thus, the small $\beta$ regime is the only one in which one obtains inflation in the $V_0 = 0$ case.

In the $V_0 \neq 0$ case one finds
\begin{equation}
\phi(N)-\phi_0=\mp\sqrt{\frac{3\beta}{2M_p^2}}\text{arcsinh}\left(\frac{\dot{\phi}_0 e^{-3\beta N}}{\sqrt{2\beta V_0}}\right) \, .
\end{equation}
Thus, one finds
\begin{eqnarray}
e^{-6\beta N}=\frac{2\beta V_0}{\dot{\phi}_0^2}\sinh^2\left(\mp\sqrt{\frac{3\beta}{2M_p^2}}(\phi-\phi_0)\right) \, ,
\end{eqnarray}
which from the Eq. (\ref{eq:VCold}) gives
\begin{eqnarray}\label{eq:VCold22}
V&=&V_0+(1-\beta)V_0\sinh^2\left(\mp\sqrt{\frac{3\beta}{2M_p^2}}(\phi-\phi_0)\right)\\
\nonumber
&=&\frac{V_0}{2}\left(1+\beta+(1-\beta)\cosh\left(\sqrt{6\beta}\frac{\phi-\phi_0}{M_p}\right)\right) \, .
\end{eqnarray}
{Note that this solution is in fact a superposition of a constant term and $\pm$ solutions from the Eq. (\ref{eq:VnoV0Cold}) which agrees with Eq. (11) in Ref.\cite{Motohashi:2014ppa}.}
\\*

The other option to obtain a potential that would satisfy the constant roll condition is to note, that
\begin{equation}
-2M_p^2\dot{H} = -2M_p^2H_\phi \dot{\phi} \quad \Rightarrow \quad \dot{\phi} = -2M_p^2 H_\phi \, , \label{eq:Fried2CRCold}
\end{equation}
where $H_\phi = \frac{dH}{d\phi}$. Implementing the Eq. (\ref{eq:Fried2CRCold}) into (\ref{eq:CR}) one finds
\begin{equation}\label{Hubble}
H_{\phi\phi} = \frac{3\beta}{2M_p^2} H \qquad \Rightarrow \qquad H \propto \exp\left(\pm \sqrt{\frac{3\beta}{2}}\frac{\phi}{M_p}\right) \, ,
\end{equation}
which is fully consistent with the Eq. (\ref{eq:VnoV0Cold}). {The form of the Hubble parameter in Eq.(\ref{Hubble}) is also presented in Eq.(10) of Ref.\cite{Motohashi:2014ppa}.}


\section{Warm constant roll inflation} \label{sec:Warm}

In this section, we follow the procedure introduced in Sec. \ref{sec:Cold} to solve the Einstein equations for the warm constant roll inflation. We assume that the inflaton can dissipate into relativistic degrees of freedom, which leads to the non-zero temperature of the Universe. {In this framework we want to obtain a constant-roll solution.}  Note that for the warm inflation the Eqs. (\ref{eq:Fried1},\ref{eq:Fried2}) still hold. Furthermore, since we require constant roll inflation, Eqs. (\ref{eq:CR},\ref{eq:phipCR}) are valid as well. The dissipation between energy densities of inflaton and radiation modifies continuity equations, which gives
\begin{eqnarray}
\ddot{\phi} + (3H+\Gamma)\dot{\phi} + V_\phi = \ddot{\phi} + 3H(1+Q)\dot{\phi} + V_\phi  = 0 \, , \label{eq:EOMWarm} \\
\dot{\rho}_r + 4H \rho_r = \dot{\phi}^2\Gamma = 3HQ\dot{\phi}^2 \, , \label{eq:ContRad}
\end{eqnarray}
where $3HQ = \Gamma$ and $\Gamma$ is a dissipation coefficient. Warm inflation with big cosmic friction enables slow evolution of the inflaton and quasi de Sitter Universe even if the potential itself does not meet the requirements of the slow-roll approximation. This effect relaxes the constraints of inflationary potentials, enabling more theories to be possibly consistent with the data. Another motivation for warm inflation comes from the swampland conjecture. In Ref. \cite{Motaharfar:2018zyb} it was proven that inflation may be consistent with the swampland for $Q \gg 1$ ( in $Q=contant$ case). 

{Just like in Sec. \ref{sec:Cold}, we require the constant-roll evolution of the field, which means that Eqs. \eqref{eq:CR} and \eqref{eq:phipCR} are still valid. $\beta\equiv -\ddot{\phi}/(3H\dot{\phi})$ remains constant and it still parametrizes the deviation from the slow-roll approximation, with $\beta\ll1$ being equivalent to the slow-roll case. Our goal is to obtain the evolution of the field and the form of the potential for given $\beta$ and $Q$.}
Note that the Eq. (\ref{eq:Fried2CRCold}) cannot be used in the presence of radiation, since the RHS of the second Friedmann equation contains (in addition to the $\dot{\phi}^2$ term) radiation term $4\rho_r/3$. {Therefore, we will follow the logic presented in Eqs. (\ref{eq:phiofN}-\ref{eq:VCold22}) and we will solve Friedmann equations using e-folds as a time variable. Since we assume the constant-roll, one can still use the Eq. \eqref{eq:phipCR}, which gives the RHS of the Eq. \eqref{eq:ContRad} to be $3HQ\dot{\phi}^2 = 3HQ\dot{\phi}_0^2e^{-6\beta N}$. The} Eq. (\ref{eq:ContRad}) can be solved analytically for $Q=const$ {and therefore} in this work we will assume $Q = const$, leaving the $Q=Q(\phi,\dot{\phi},T)$ case for future analysis. From the Eq. (\ref{eq:phipCR}) one finds
\begin{equation}
\rho_r = \rho_{r0} a^{-4} + \frac{3Q\dot{\phi}_0^2}{2(2-3\beta)}a^{-6\beta} \, , \label{eq:RhoR}
\end{equation}
where the $\rho_{r0}$ term is a radiation energy density without dissipation. This solution {can be analyzed in two limits. }
\begin{itemize}
\item[a)] {For $\beta\ll 1$ it is fully consistent with a standard slow-roll  warm inflation. In such a case} one finds the almost constant value of $\rho_r$. 

\item[b)] In the case of significant deviation from the slow-roll, one obtains strong, exponential suppression of $\rho_r$. This leads to the inflationary scenario, {with mixed features of both, cold and warm inflation. The cosmic friction is significantly increased (which is a characteristic feature of warm inflation), but the temperature is very close to the absolute zero, as in the case of cold inflation. The $\beta>2/3$ case seems to be especially interesting, since in the Eq. \eqref{eq:RhoR} the term proportional to $Q$ becomes negative. Thus, one could wrongly conclude that for $\beta>2/3$ the radiation obtains smaller energy density (and therefore lower temperature) than in the case of cold inflation, even though the dissipation of energy density goes exclusively from the inflaton to radiation. In fact, warm inflation \emph{always} leads to a slower decrease of the temperature comparing to the cold inflationary scenario. In order to see that let as consider $\rho_r$ from the Eq. \eqref{eq:RhoR} and unsourced radiation $\tilde{\rho}_r = \tilde{\rho}_{r0}a^{-4}$. The $\rho_r/\tilde{\rho}_r$ ratio will \emph{always} grow, since 
\begin{equation}
\frac{d}{dt}\left(\frac{\rho_r}{\tilde{\rho}_{r0}a^{-4}}\right) = 3HQ\frac{\dot{\phi}_0^2}{\tilde{\rho}_{r0}}a^{4-6\beta} > 0 \qquad \text{for all $\beta$ \, .}
\end{equation}
Since the radiation contains a term, which can be negative, one could in principle consider $N$ small enough that would result in negative radiation energy density, which is unphysical. In order to avoid that let us restrict our analysis to $N> -\frac{1}{2(3\beta-2)}\log(2(3\beta-2)\rho_{r0}/(3Q\dot{\phi}_0^2))$, which secures $\rho_{r}>0$. Again, this constraint on $N$ is necessary only for $\beta>2/3$, since for $\beta<2/3$ one finds $\rho_r>0$ for all $N$.}
\end{itemize}

One can employ the Eq. (\ref{eq:RhoR}) to solve the Eq. (\ref{eq:Fried2}), which takes the form of
\begin{equation}
-M_p^2\frac{d}{dN}(H^2) = \dot{\phi}_0^2\frac{2+2Q-3\beta}{2-3\beta} e^{-6\beta N} + \frac{4}{3}\rho_{r0}e^{-4N} \, . \label{eq:Fried2Warm}
\end{equation}
By integrating both hand sides with respect to $N$ one obtains the first Friedmann equation 
\begin{equation}
3M_p^2H^2 = \rho = \frac{\dot{\phi}_0^2}{2\beta}\frac{2+2Q-3\beta}{2-3\beta} e^{-6\beta N} + \rho_{r0}e^{-4N} + V_0 \, . \label{eq:Fried1SolWarm}
\end{equation}
For $Q>0$ the $a^{-6\beta}$ term is positive for $\beta < 2/3$ or for $\beta > 2(1+Q)/3$. For $\rho_{r0} = V_0 = 0$ one finds $\epsilon = 3\beta$, which puts the condition $\beta \ll 1$ for the inflation with $V_0 = 0$. This is fully consistent with the cold inflation case from the Eq. (\ref{eq:EpsilonCold}). Like in the constant roll cold inflation one can extract the potential $V$ from the energy density using Eqs. (\ref{eq:Fried1},\ref{eq:Fried1SolWarm}), which gives
\begin{equation}
V =V_0 + \frac{\dot{\phi}_0^2}{2\beta}(1+Q-\beta)a^{-6\beta} \, . \label{eq:VWarm}
\end{equation}
We want to emphasize that in the case of the warm inflationary scenario one obtains a wider range of allowed values of $\beta$, namely $\beta \leq 1+Q$. The second way of obtaining $V$ is via the effective equation of motion of $\phi$, namely
\begin{equation}
 3H(1+Q-\beta)\dot{\phi} + V_\phi = 0 \, . \label{eq:CRWarm}
\end{equation}
By changing the derivatives into $\frac{d}{dN}$, which follows the procedure described in the Eq. (\ref{eq:VCold2}), one restores the Eq. (\ref{eq:VWarm}). Note that for $\beta = 1+Q$ one finds $V = V_0$. This case is significantly different from the ``cold'' constant roll cold inflation. Since $Q$ may be in principle much bigger than one (for $\Gamma \gg 3H$) one may obtain $\beta \gg 1$, which increases the deviation from the slow-roll regime.

{The Eq. \eqref{eq:VWarm} gives the general form of $V(N)$.} In order to obtain $V = V(\phi)$ one needs to {find $\phi(N)$ and reverse it to $N(\phi)$. Solving the Eq. (\ref{eq:phiNCondition})} gives
\begin{equation}
\phi(N) = \pm \sqrt{3}\dot{\phi}_0 M_p \int \frac{e^{-3\beta N} dN}{\sqrt{\frac{\dot{\phi}_0^2}{2\beta}\frac{2+2Q-3\beta}{2-3\beta} e^{-6\beta N} + \rho_{r0}e^{-4N} + V_0}} \, . \label{eq:phiNWarm}
\end{equation}
For general value of $\beta$ the solution for $\phi(N)$ cannot be founded. Nevertheless the approximate solution can be founded in certain regimes. For instance, the denominator of the Eq. (\ref{eq:phiNWarm}) simplifies, if the term $\frac{\dot{\phi}_0^2}{2\beta}\frac{2+2Q-3\beta}{2-3\beta} e^{-6\beta N}$ is much bigger than $V_0$ and $\rho_{r0}e^{-4N}$. This can be satisfied for sufficiently big $a$ (i.e. for sufficiently late times) for $\beta <2/3$ and $V_0 = 0$ or for $\beta > 2/3$ for sufficiently small $a$. In such a case one finds
\begin{equation}
\phi = \pm M_p \sqrt{\frac{6\beta (2-3\beta)}{2+2Q-3\beta}}N + \phi_0 \qquad \Rightarrow \qquad N = \pm \frac{\phi - \phi_0}{M_p}\sqrt{\frac{6\beta (2-3\beta)}{2+2Q-3\beta}} \, . \label{eq:phiLinearWarm}
\end{equation}
The Eq. (\ref{eq:phiLinearCold}) gives the same result as (\ref{eq:phiLinearWarm}) if one redefines $\beta$ in (\ref{eq:phiLinearCold}) into $\beta(2-3\beta)/(2+2Q-3\beta)$. The scalar potential in the (\ref{eq:phiLinearWarm}) is equal to
\begin{equation}
V = \frac{1+Q-\beta}{2\beta} \dot{\phi}_0^2\exp\left(\mp \sqrt{\frac{6\beta (2-3\beta)}{2+2Q-3\beta}}\frac{\phi-\phi_0}{M_p}\right) \, . \label{eq:VnoV0Warm}
\end{equation}
One can also solve the Eq. (\ref{eq:phiNWarm}) analytically by assuming, $\rho_{r0} = 0$, which means that there is no radiation other than the one produced by dissipation. This case can be especially realistic for $\beta \ll1$, for which the $a^{-6\beta}$ term in the Eq. (\ref{eq:RhoR}) redshifts much slower than the $a^{-4}$ term. In such a case one finds:
\begin{eqnarray}
\phi-\phi_0=\mp\sqrt{\frac{2M_p^2(2-3\beta)}{3\beta(2+2Q-3\beta)}}\text{arcsinh}\left(\sqrt{\frac{\dot{\phi}_0^2(2+2Q-3\beta)}{2\beta V_0(2-3\beta)}}e^{-3\beta N}\right) \, ,
\end{eqnarray}
\\*
which leads to
\begin{eqnarray}
e^{-6\beta N}=\frac{2\beta V_0(2-3\beta)}{\dot{\phi}_0^2(2+2Q-3\beta)}\sinh^2\left(\mp\sqrt{\frac{3\beta(2+2Q-3\beta)}{2M_p^2(2-3\beta)}}(\phi-\phi_0)\right) \, .
\end{eqnarray}
Using the above equation and Eq. (\ref{eq:VWarm}), one can find the potential 
\begin{eqnarray}
V=V_0\left(1+\frac{(1+Q-\beta)(2-3\beta)}{(2-3\beta+2Q)}\sinh^2\left(\mp\sqrt{\frac{3\beta(2+2Q-3\beta)}{2M_p^2(2-3\beta)}}(\phi-\phi_0)\right) \right) \, ,
\end{eqnarray}
which agrees with cold result in the limit $Q=0$ Eq. (\ref{eq:VCold22}) and Eq. (19) in Ref.\cite{Motohashi:2014ppa}. 

The analysis we present is valid for $\beta \neq 2/3$, since the energy density of radiation has a pole in $3\beta-2$. For $\beta = 2/3$ one finds 
\begin{equation}
\rho_r = \left(\rho_{r0} + 3Q\dot{\phi}_0^2 \log a\right)a^{-4} = \left(\rho_{r0} + 3Q\dot{\phi}_0^2 N\right) e^{-4N} \, ,
\end{equation}
which means that
\begin{equation}
3M_p^2H^2 = \rho = \left(\frac{3}{4}\dot{\phi}_0^2 \left(1 + Q(4N+1) \right) + \rho_{r0} \right) e^{-4N} + V_0 \, . \label{eq:Fried1SolWarm2to3}
\end{equation}
The energy density is always positive for $Q>0$. For $\beta = 2/3$ the $\phi = \phi(N)$ relation takes the form of
\begin{equation}
\phi(N) = \pm \sqrt{3}\dot{\phi}_0 M_p \int \frac{e^{-2 N} dN}{\sqrt{ \left(\frac{3}{4}\dot{\phi}_0^2 \left(1 + Q(4N+1) \right) + \rho_{r0} \right) e^{-4N} + V_0}} \, . \label{eq:phiN2to3Warm}
\end{equation}
For $V_0 = 0$ the Eq. (\ref{eq:phiN2to3Warm}) gives
\begin{equation}
\phi = \pm \frac{2M_p}{\sqrt{3}Q\dot{\phi}_0}\sqrt{\frac{3}{4}\dot{\phi}_0^2 \left(1 + Q(4N+1) \right) + \rho_{r0}} + \phi_0 \, . \label{eq:phiNSol2to3}
\end{equation}
Note that in the $Q \to 0$ limit one finds $\phi \propto N$, which corresponds to the cold constant roll scenario with $V_0 = 0$. From (\ref{eq:VWarm},\ref{eq:phiNSol2to3}) one finds
\begin{equation}
V \propto \exp\left(-Q\frac{(\phi-\phi_0)^2}{M_p^2}\right) \, .
\end{equation}
In the cold inflationary Universe inflation around the maximum of the Gaussian potential generates way to small $n_s$ to be consistent with the Planck data. We leave the issue of the observational predictions of the warm Gaussian inflation for future analysis. 
\\* 

Note that for $Q>0$ and $\beta > 2/3$ the $a^{-6\beta}$ term in the Eq. (\ref{eq:RhoR}) becomes negative. {Nevertheless, the total energy density is always positive, since both $\rho_\phi$ and $\rho_r$ are positive by definition.} 


\section{Evolution of primordial inhomogeneities} \label{sec:Pert}

\subsection{The $V_0 = 0$ case}

The simplest regime in which one can investigate the evolution of inhomogeneities is the small $\beta$ limit, for which one does not deviate significantly from the slow-roll evolution. In such a case the energy density of the Universe should be dominated by the energy density of $\phi$. The evolution of radiation should be determined by the term that comes from the dissipation (i.e. the $a^{-6\beta}$ term). One can also assume that $V_0 = 0$, which follows from the assumption that the $V(\phi)$ is a full potential of the theory that describes the evolution of the field during all of its evolution. In such a case one finds $\epsilon =-\frac{\dot{H}}{H^2}= 3\beta$ and $\frac{V_{\phi\phi}}{H^2} = -18(1+Q)\beta$, which can be {employed in the equation of motion of perturbations of the inflaton}. We want to emphasize that taking $V_0 = 0$ makes the model effectively a warm inflationary power-law inflation. The ``cold'' version of this theory has been investigated in e.g. Eq. (15) in \cite{Motohashi:2014ppa}.

In the warm inflationary scenario the evolution of fluctuation modes is divided into three regimes: 1) thermal noise, 2) expansion and 3) curvature fluctuations \cite{Hall:2003zp}. Freeze out is the transition between regimes 1) and 2) and the horizon crossing is the transition between regimes 2) and 3). Thermal noise regime is studied by Schwinger-Keldeysh approach to non-equilibrium field theory \cite{Hall:2003zp} where the evolution of the inflaton field is modified as a Langevin equation:
\begin{eqnarray}
-\square\phi(x,t)+\Gamma\dot{\phi}+V_{\phi}=(2\Gamma_{eff}T)^{\frac{1}{2}}\xi(x,t)
\end{eqnarray} 
where $\square$ is the space-time Laplacian, $\Gamma_{eff} = H(1+Q)$ and $\xi$ is added stochastic term {with approximately Gaussian probability distribution}\cite{Berera:2007qm,Graham:2009bf,Hall:2003zp} {and two-point} correlation function:
\begin{eqnarray}
\langle\xi(x,t)\xi(x',t')\rangle=\delta^3(x-x')\delta(t-t')
\end{eqnarray}
Using equivalence principle one can express the Langevin equation in curved FRW space-time for perturbation of inflaton field $\delta\phi(x,t)$ where
\begin{eqnarray}\label{scalar}
\phi(x,t)=\phi(t)+\delta\phi(x,t)
\end{eqnarray}
and
\begin{eqnarray}
\langle\xi(x,t)\xi(x',t')\rangle=a^{-3}(2\pi)^2\delta^3(x-x')\delta(t-t')
\end{eqnarray}
which is the new definition of the Gaussian condition of noise term in the FRW space-time. $\delta\phi$ is a linear response due to small perturbation noise $\xi$, which is a function of space-time, and $\phi(t)$ is a background part of the inflaton field. Under the assumption $T>H$, the perturbation part of Langevin equation in an expanding FLRW universe during warm inflation can be presented after Fourier transforms \cite{Graham:2009bf,Hall:2003zp}:
\begin{eqnarray}\label{Perturbed Langevin}
\delta\ddot{\phi}(\textbf{k},t)+(3H+\Gamma)\delta\dot{\phi}(\textbf{k},t)+V_{\phi\phi}\delta\phi(\textbf{k},t)+k^2a^{-2}\delta\phi(\textbf{k},t)=(2\Gamma_{eff}T)^{\frac{1}{2}}\xi(\textbf{k},t)
\end{eqnarray} 
In the thermal noise era, generation of space-time inhomogeneities due to inflaton fluctuation $\delta\phi$ can be discarded in uniform expansion rate gauge  \cite{Moss:2007cv} and sub-horizon scale where the scales are smaller than the horizon and bigger than thermal averaging scale. New time variable $z=\frac{k}{aH}$ can simplify the perturbed  Langevin equation (\ref{Perturbed Langevin}) in term of the constant-roll parameters
\begin{eqnarray}
\left(1+\frac{\dot{H}}{H^2}\right)\delta\phi''-(3Q+2)\frac{1}{z}\left(1+\frac{\dot{H}}{H^2}\right)\delta\phi'+\left(1+\frac{\dot{H}}{H^2}\right)\left(\frac{\dot{H}}{H^2}\right)'\delta\phi'\nonumber\\
+\frac{V_{\phi\phi}}{H^2}z^{-2}\delta\phi+\delta\phi= (2\Gamma_{eff}T)^{\frac{1}{2}}\left(\frac{a}{k}\right)^2\xi(\textbf{k},t) \, , \label{eq:EOMz}
\end{eqnarray}    
 where $' = \frac{d}{d z}$. In the $\beta \ll 1$ regime the Eq. \eqref{eq:EOMz} simplifies into
 \begin{equation}
(1-3\beta)\delta\phi'' - (3Q+2)(1-3\beta)\frac{1}{z}\delta\phi' -18(1+Q)\beta z^{-2}\delta\phi+\delta\phi = (2\Gamma_{eff}T)^{\frac{1}{2}}\left(\frac{a}{k}\right)^2\xi(\textbf{k},t) \, . \label{Pert1}
\end{equation}
Note that this equation takes form of the equation of motion with a thermal noise. The Eq. \eqref{Pert1} refers to the Eq. (54) from the Ref. \cite{Graham:2009bf} with the substitution  $z\rightarrow z\sqrt{1-3\beta}$: 
\begin{equation}\label{Pert}
\frac{d^2\delta\phi}{dz^2} - (3Q+2)\frac{1}{z}\frac{d\delta\phi}{dz} -\frac{18(1+Q)\beta}{1-3\beta} z^{-2}\delta\phi+\delta\phi = (2\Gamma_{eff}T)^{\frac{1}{2}}\left(\frac{a}{k}\right)^2\xi(\textbf{k},t) \, .
\end{equation}
{We have considered $\Gamma \propto H$, which mean that $\Gamma$ does not depend on temperature. Therefore} we have assumed that the back reaction of perturbations of radiation on $\Gamma$ is negligible. This issue is investigated in the Ref.\cite{Graham:2009bf} in the context of the slow-roll inflation. The above equation without source is comparable with Eq. (\ref{GB}) in the appendix, where $\nu=\frac{3}{2}(Q+1),$~$\gamma=1,\beta=1$, $n\simeq\pm (\nu+3c)$ and  $c=\frac{2(1+Q)\beta}{1-3\beta} \simeq 2\beta(1+Q)$. First of all, let us study the $n=\nu+3c$ case. As we will show, the negative $n$ scenario is ruled out, since it predicts a value of spectral index, which is inconsistent with data constraints.
 
 The solution of this equation is found from the Green function technique as
\begin{eqnarray}\label{Pert2}
\delta\phi=\int_{z}^{\infty}G^{c}(z,z')(z')^{1-2\nu}(2\Gamma_{eff}T)^{\frac{1}{2}}\hat{\xi}(z')dz' \, ,
\end{eqnarray} 
where the Green function of perturbation equation (\ref{Pert}) is presented by
\begin{eqnarray}\label{Green}
G^{c}(z,z')=\frac{\pi}{2}z^{\nu}z^{'\nu}\left(J_{\nu+3c}(z)Y_{\nu+3c}(z')-J_{\nu+3c}(z')Y_{\nu+3c}(z)\right)
\end{eqnarray}
and $\nu=\frac{3}{2}(1+Q)$. The power spectrum of inflation is extracted from the two-point correlation function of inflaton field perturbations
 \begin{eqnarray}\label{two-point}
 \langle\delta\phi(k,z)\delta\phi(k',z)\rangle =P_{\phi}(2\pi k)^3\delta^3(k+k') \, .
 \end{eqnarray}
One can find the power spectrum of the model from Eqs.(\ref{Pert2}),(\ref{Green}) and (\ref{two-point}) presented by \cite{Hall:2003zp,Graham:2009bf}
 \begin{eqnarray}
 P_{\phi}(k,z)=(2\Gamma_{eff}T)\int_z^{\infty}dz'G^{c}(z,z')z'^{2-4\nu} \, .
 \end{eqnarray}
In super-horizon and thermal noise part of the evolution of perturbations ($z\ll 1$), which the constant-roll is important, the power spectrum is simplified as
 \begin{eqnarray}
 P_{\phi}=(2\Gamma T)\frac{\Gamma_R(c+\frac{3}{2})}{\Gamma_{R}(\frac{3}{2})}\left(\frac{2\nu}{z^2}\right)^{3c}\sqrt{\frac{\pi}{32\nu}} \, ,
 \end{eqnarray}
where the $\Gamma_R$ is gamma function. One can re-do these calculations for the negative $n$, which gives
\begin{equation}
P_\phi\propto z^{2(2\nu+3c)} \, .
\end{equation}
Note that in the case of $n < 0$ the spectrum is blue-tilted. Using small spatial gradient expansion \cite{Salopek:1990jq}, one can define the curvature perturbation $\xi$ which is conserved even beyond the linear perturbation theory \cite{Graham:2009bf,Sasaki:1995aw,Lyth:2004gb,Lyth:2005fi}
 \begin{eqnarray}
 \xi=\frac{1}{2}\ln(1+2\varphi)+\frac{1}{3}\int\frac{d\rho}{\rho+P} \, ,
\end{eqnarray} 
where $\varphi$ is the spatial part of the curvature perturbation. In high dissipative warm constant-roll case, $\xi$ takes the form
\begin{eqnarray}
\xi=\frac{1}{2}\ln(1+\varphi)+\frac{\frac{3}{2}Q(1+\beta)+1}{1+Q}\int\frac{H}{\dot{\phi}}d\phi \, .
\end{eqnarray}
 In uniform gauge $\varphi=0$ where $\xi(\phi(x,t))$ is a function of scalar field (\ref{scalar}) one finds 
 \begin{eqnarray}
 \xi &=& \frac{d\xi}{d\phi}\delta\phi \, ,\\
 \nonumber
 P_{\xi} &=& \left(\frac{Q(1+\frac{3}{2}\beta)+1}{1+Q}\frac{H}{\dot{\phi}}\right)^2P_{\phi} \, .
\end{eqnarray} 
The spectral index is presented by
\begin{eqnarray}
n_s-1=\frac{d\ln P_{\xi}}{d\ln k}=-6c=-\frac{12(1+Q)\beta}{1-3\beta} \simeq -12(1+Q)\beta \, .\label{eq:nsSmall}
\end{eqnarray}
For the allowed $2\sigma$ region of the spectral index $n_s$ from Planck 2018 results \cite{Akrami:2018odb} we find  {$\beta (1+Q) \in (0.0022,0.0034)$, which is consistent with initial assumption} $\beta\ll1$. {On the other hand, one can consider the second branch of the solution of Eq. \eqref{Pert}, which is} {$n=-(\nu+3c)$, (see Eqs. (\ref{Pert},\ref{GB})). Then, the spectral index is equal to $n_s-1=2(2\nu+3c)=6(1+Q)(2\beta+1)$. For small $\beta$ one finds $n_s \simeq 6(1+Q)$, which is highly disfavoured by the Planck data.}

Tensor perturbation of space-time metric is not affected by thermal bath during warm inflation \cite{Taylor:2000ze}. Therefore one can use the standard definition of cold inflation tensor power-spectrum at the horizon crossing $P_T=\frac{8}{M_p^2}(\frac{H}{2\pi})^2$. Another important perturbation parameter which can be constrained by observational data is the tensor-to-scalar ratio 
\begin{eqnarray}\label{Tensor}
r=\frac{P_T}{P_{\xi}}=\frac{1}{\pi^2M_p^2}\frac{\Gamma_R(\frac{3}{2})}{\Gamma_R(c+\frac{3}{2})}\frac{(32\nu)^{\frac{1}{2}}}{(2\nu)^{3c}\sqrt{\pi}}\left(\frac{1+Q}{1+Q(1+\frac{3}{2}\beta)}\right)^2\frac{\dot{\phi}^2}{3(1+Q)HT} \, .
\end{eqnarray}
{Note, how big values of $Q$ lead to a $Q^{-1}$  suppression of $r$. This effect enables us to fit the Planck data, by fixing $\beta(1+Q)$ with sufficiently big $Q$.} \

\begin{figure}
\begin{center}
\includegraphics[scale=0.45]{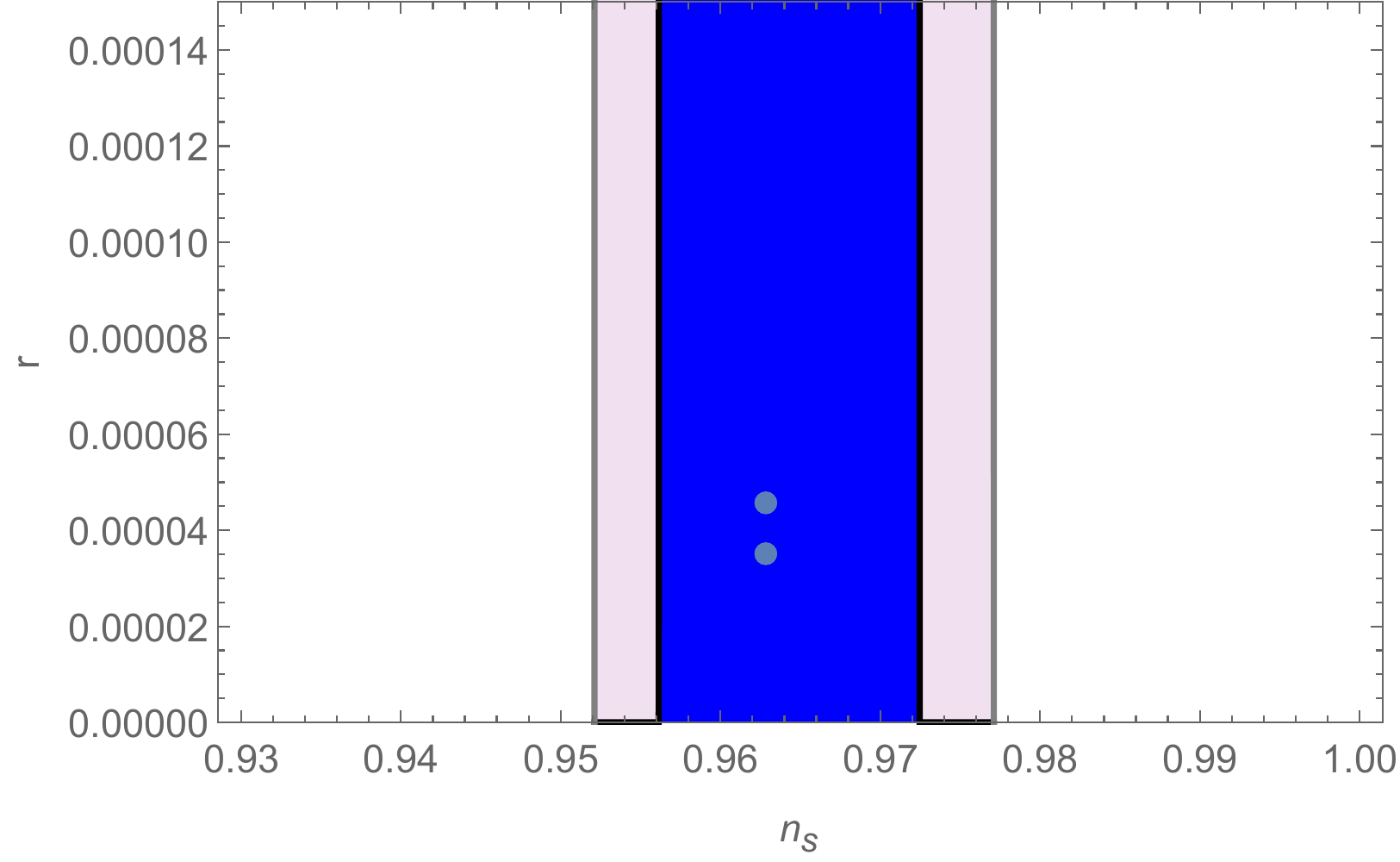}
\end{center}
\caption{$(n_s,r)$ of our model for two cases $N=60$ (lower point) and $N=50$ (upper point) is in $1-\sigma$ confidence level of $r-n_s$ Planck results \cite{Akrami:2018odb}.($\beta=10^{-6},~Q=2.8\times 10^2,\dot{\phi}_0=\frac{\sqrt{C_{\gamma}}M_p^2}{(0.13)^2}$)}
\label{r-n}
\end{figure}

 In Fig. \ref{r-n} the results of our model from Eqs. \eqref{eq:nsSmall} and \eqref{Tensor} are compared with observational data. There is a viable choice for phase space of the model parameters ($Q,\beta$) which lies within $1\sigma$ confidence level of $n_s-r$ Planck data results \cite{Akrami:2018odb}.

\subsection{The $V_0 \neq 0$ case} 

\begin{figure}
\begin{center}
\includegraphics[scale=0.55]{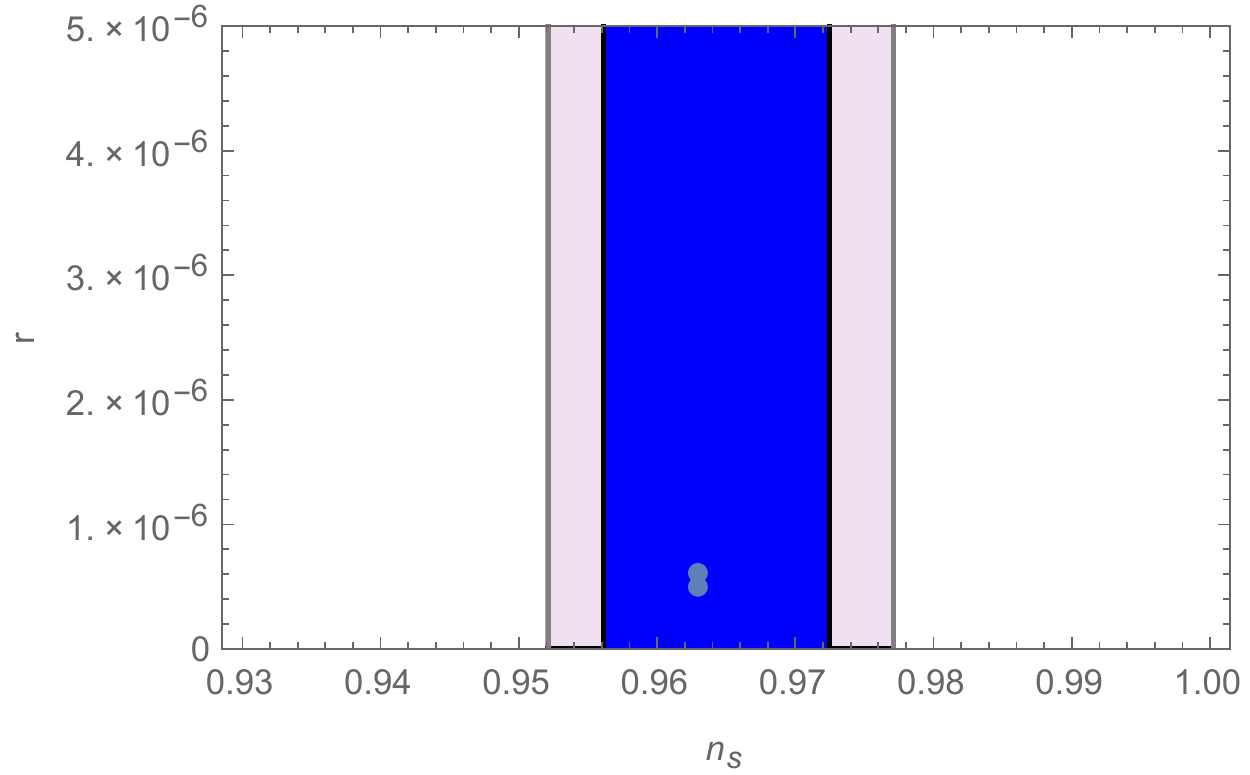}
\end{center}
\caption{$(r,n_s)$ of our model for two cases $N=60$ (lower point) and $N=50$ (upper point) is in $1-\sigma$ confidence level of $r-n_s$ Planck results \cite{Akrami:2018odb}.($\beta=2.5\times 10^{-5},~Q= 2\times10^{2},\dot{\phi}_0^{1.5}=6\times 10^{-5}\sqrt{V_0}M_p$). The value of $Q$ has been {chosen} to obtain correct value of $n_s$.}
\label{r-n1}
\end{figure}

 If we chose the constant potential $V_0$ as a dominant part of the energy density during the constant-roll warm inflation, the above procedure of perturbation part can be repeated and the more generic values of $\beta$ can be considered. In this regime the slow-roll parameter $\epsilon$ is very small and it evolves like $\epsilon \propto a^{-6\beta}$. In such a case the $\epsilon$ parameter can be fully neglected in Eq. \eqref{eq:EOMz}, which gives
\begin{equation}\label{Pert3}
\frac{d^2\delta\phi}{dz^2} - (3Q+2)\frac{1}{z}\frac{d\delta\phi}{dz} -9(1+Q-\beta)\beta z^{-2}\delta\phi+\delta\phi = (2\Gamma_{eff}T)^{\frac{1}{2}}\left(\frac{a}{k}\right)^2\xi(\textbf{k},t) \, .
\end{equation}
The solution of this equation is also given by (\ref{Green}) with new parameter $c=\beta(1+Q-\beta)$. We can continue another steps similar to previous case to find spectral index $n_s$
\begin{eqnarray}
n_s-1=\frac{d\ln P_{\xi}}{d\ln k}=-6c=-6\beta(1+Q-\beta) \, . \label{eq:nsanybeta}
\end{eqnarray}
One can obtain $n_s \simeq 1$ in two limits: for $\beta \ll 1$ or for $\beta \simeq 1+Q$. The first case is similar to the one discussed in the previous part of this section. The latter one corresponds to a strong deviation from the slow-roll approximation. In the small $\beta$ limit one can use the constraints on $n_s$ to limit the allowed values of $\beta$ into $\beta (1+Q) \in (0.0044,0.0068)$. The tensor-to-scalar ratio of this case is presented by Eq. (\ref{Tensor}). In Fig. (\ref{r-n1}) we show the consistency of our model with the Planck observational data.

For the $n=-(\nu+3c)$ solution (see Eqs.(\ref{Pert},\ref{GB}), the spectral index is
\begin{equation}
n_s-1=2(2\nu+3c)=6\beta (1+Q-\beta)+6(1+Q) \, .
\end{equation}
 Such values of $n_s$ are clearly ruled out by the Planck observational data, which requires $1-n_s \ll 1$. In fact one could fine tune the relation between $\beta$ and $Q$, for which the spectrum would remain flat. However, in such a case one finds $\beta > 1+Q$, which generates $V < 0$ or at least $V_N > 0$. It means that the only scale-invariant spectrum may be generated in the \eqref{eq:nsanybeta} scenario. 

One of the most interesting cases of the constant-roll inflation is $\beta \simeq 1+Q$, for which one obtains the strongest possible deviation from the slow-roll approximation, while still maintaining the de Sitter evolution of the Universe. In such a case the RHS of the Eq. (\ref{eq:Fried2Warm}) contains 2 terms: (1) proportional to $a^{-6(1+Q)}$, which comes from the kinetic term of the inflaton field as well as from the radiation-induced by the source term in the Eq. (\ref{eq:ContRad}), (2) the $a^{-4}$ term, which comes from the pre-existing radiation. In both cases one finds $\eta = -3(1+Q)$ In a case of an $a^{-6(1+Q)}$ domination one finds
\begin{equation}
\epsilon \propto \frac{\dot{\phi}^2}{V_0} \propto a^{-6(1+Q)} \, , 
\end{equation}
while for the $a^{-4}$ term domination one obtains
\begin{equation}
\epsilon \propto \frac{\rho_r}{V_0} \propto a^{-4} \, .
\end{equation}
In the case of the $a^{-6\beta}$ term domination in $\dot{H}$, the influence of radiation can be neglected in the analysis. Thus, one can analyze the evolution of primordial inhomogeneities in the framework of cold inflation. In such a case one finds \cite{Motohashi:2014ppa}
\begin{equation}
\zeta_k \simeq A_k + B_k a^{3(2\beta-1)} =  A_k + B_k a^{3(1+2Q)} \, ,
\end{equation}
where $\zeta_k$ is a Fourier mode of a curvature perturbation and $A_k$, $B_k$ are constants. For $\beta < 1/2$ one finds $\zeta_k \to A_k$ for $a \to \infty$, which is equivalent to the super-horizon freeze-out of curvature perturbations in the standard slow-roll inflation. For $\beta > 1/2$ the super-horizon modes tend to grow in time, which leads to the amplification of inhomogeneities. This mechanism may be used in order to generate primordial black holes{(BH), which in the context of cold constant-roll inflation is already discussed in \cite{Motohashi:2019rhu}}.  This mechanism may be much more efficient than in the case of the cold constant roll inflation, since for $Q > 0$ the $\beta$ is allowed to take values bigger than 1. This could decrease fine tuning on the process of the primordial BH production. We want to investigate this issue in our further work.

\section{Conclusions} \label{sec:Conclusions}

In this paper, the constant roll evolution of the warm inflationary Universe has been investigated. {Throughout the whole work we have assumed that the inflaton field satisfies the constant roll equation of motion, namely $\ddot{\phi} + 3\beta H \dot{\phi} = 0$, where $\beta$ is a constant.}  In the Sec. \ref{sec:Cold} {we assume that the inflaton is \emph{not} coupled to any other fields and therefore inflation is cold. Sec. \ref{sec:Cold} does not contain a new constant-roll inflationary model, but it contains a novel approach to reconstructing the inflationary potential that would secure the constant roll evolution of the field}. Instead of using a field as a variable, we use the number of e-folds ($N$), which appears to be a useful tool in warm inflation. An analytical solution for Hubble parameter $H$, scalar field $\phi$ and its potential $V(\phi)$ as a function of $N$ have been founded. Finally, it was shown that this approach is fully consistent with the analytical solutions obtained so far in constant-roll inflation.
\\*

In Sec. \ref{sec:Warm} this analysis was extended to the warm inflationary scenario, in which the inflaton dissipates towards relativistic degrees of freedom with a dissipation coefficient $\Gamma$. In the simplest case of $\Gamma =  3 Q H$ (where $Q = const$) a series of analytical solutions for $H$, $\phi$, $V(\phi)$ and energy density of radiation $\rho_r$ have been obtained. It was shown that $\rho_r$ contains an $a^{-4}$ term, which comes from the pre-inflationary radiation, as well as the $a^{-6\beta}$ term, which comes from the dissipation and redshifts like a kinetic term of the inflaton field. For $\beta > 2/3$ the $a^{-6\beta}$ term is negative. Nevertheless, the energy density of radiation still decays slower than in the case of cold inflation.

One of the simplest solutions for the constant roll warm inflation founded by us is the $V = const$, $\beta = 1+Q$ scenario. In such a case one finds $\eta = -3\beta$ (where $\beta > 1$) and $\epsilon \propto a^{-6\beta}$, which is the case of the strongest deviation from the slow-roll approximation. Another interesting case is $\beta = 2/3$, for which one finds $\rho_r/\dot{\phi}^2 \sim N$. This is the only solution founded in this paper, for which radiation significantly dominates the right-hand side of the second Friedmann equation. 
\\*

In Sec. \ref{sec:Pert} the evolution of primordial inhomogeneities have been analyzed in two cases. {For} $\beta, \ll 1$, which denotes small deviation from slow-roll, the power spectrum of curvature perturbations with a small deviation from scale invariance has been founded. From observational bounds on $n_s$ we have obtained allowed range of $\beta \in (0.0022,0.0034)$. Second of all, the more general case has been considered for $\epsilon \propto a^{-6\beta}$ and any value of $\beta$. One finds $n_s \simeq 1$ in two cases: for $\beta \ll 1$ and $\beta \simeq 1 + Q$. 

The $\beta = 1+Q$ case was briefly discussed, for which one finds the exponential growth of super-horizon modes of the curvature perturbation $\zeta$. The growth of the super-horizon inhomogeneities is stronger than in the case of cold constant roll inflation, due to a bigger value of $\beta$. We conclude the constant-roll warm inflation may be a highly successful inflationary theory only for $\beta \ll1$. Nevertheless, some parts of the potential may be characterized by $\beta > 2/3$, which may lead to the growth of primordial inhomogeneities. {This mechanism may be used to produce primordial BH and to decrease the fine-tuning of the part of the potential responsible for the BH production.}

\appendix 
{\section{Bessel function}
Generalized Bessel equation is presented by 
\begin{eqnarray}\label{GB}
x^2\frac{d^2y}{dx^2}+(1-2\nu)x\frac{dy}{dx}+(\beta^2\gamma^2x^{2\gamma}+(\nu^2-n^2\gamma^2))y=0 \, ,
\end{eqnarray}
which has the general solution:
\begin{eqnarray}\label{SoB}
y(x)=Ax^{\nu}J_{n}(\beta x^{\gamma})+Bx^{\nu}Y_{n}(\beta x^{\gamma}) \ ,
\end{eqnarray}
The \eqref{SoB} solution has been used in the context of primordial inhomogeneities in Sec \ref{sec:Pert}.}

\acknowledgments
V.K's research at McGill has been supported by a NSERC Discovery Grant
to Robert Brandenberger and the McGill Space Institute.
This work has been supported by the National Science Centre, Poland, under research grant DEC-2012/04/A/ST2/00099. M.A. thanks M. Malekjani and Bu-Ali Sina University for hospitality and Misao Sasaki for his comments.

\end{document}